# Design and properties of hollow antiresonant fibers for the visible and near infrared spectral range

Walter Belardi

Optoelectronics Research Centre, University of Southampton, Southampton, SO17 1BJ, UK

w.belardi@soton.ac.uk

*Abstract*—Hollow core antiresonant fibers offer new possibilities in the near infrared and visible spectral range. I show here that the great flexibility of this technology can allow the design and fabrication of hollow core optical fibers with an extended transmission bandwidth in the near infrared and with very low optical attenuation in the visible wavelength regime. A very low attenuation of 175dB/km at 480nm is reported. A modification of the design of the studied fibers is proposed in order to achieve fast-responding gas detection.

## I. Introduction

OPTICAL TRANSMISSION in hollow waveguides has been investigated since the dawn of optical fiber technology [1]. However, only the later development of Hollow Core Photonic Band-Gap Fibers (HC-PBGFs) [2] provided a viable means for efficient light guidance in air. These fibers are characterized by the presence of a central air hole surrounded by periodical layers of air holes in the cladding structure and have shown a minimum optical attenuation of 1.2dB/km in the telecommunication band [3]. However HC-PBGF performances rapidly deteriorate at shorter wavelengths $\lambda$, due to the incidence of optical scattering, which scales as $\lambda^{-3}$ [3]. To date the minimum attenuation of commercially available HC-PBGFs is about 1200dB/km at 0.532µm and 1500dB/km at 0.44µm [4].

Recent years have seen important developments of alternative typologies of Hollow Core (HC) optical fibers, based on the antiresonant effect [5]. For example, "Kagome" type HC fibers with an inverted curvature of the optical core boundary [6] have been recently demonstrated with attenuation as low as 70dB/km at 0.6µm [7].
The guidance properties of Kagome Fibers (KFs) largely depend on the thickness of the glass webs forming the cladding structure. Moreover it has been shown that the properties of KFs depend mainly on the first silica layer surrounding the optical core [8]. That is why a simplified form of HC fiber comprising a single ring of air holes in the cladding space has been proposed [9] and largely investigated. This Hollow Core Antiresonant Fiber (HC-ARF), with an inverted optical core boundary [10, 11], has been proven to have low losses in the mid-infrared wavelength regime [12]-[14], due to the reduced overlap of the fundamental-like optical mode on the fiber glass material [11]. Some modified forms of the basic design for HC-ARFs have been proposed in order to demonstrate low bending loss in the mid-infrared [14] and the possibility of reduced attenuation in the near-infrared spectral range [15]-[17].

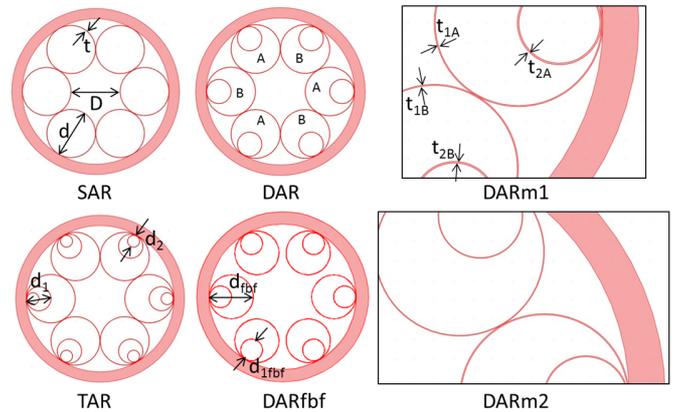

Fig. 1. Different structures of hollow core antiresonant fibers.

In this work I intend to further demonstrate the great design flexibility of HC-ARFs and some relevant applications in the near infrared and visible spectral regime. I first discuss some optical designs for obtaining an extended transmission bandwidth in the near infrared wavelength range. A theoretical bandwidth between 0.65µm and 2.5µm is numerically demonstrated and compared to some experimental results. The optical characteristics of similar HC-ARFs optimized for the near-infrared and visible spectral ranges are shown, for the first time, and their fabrication issues are discussed. A very low optical attenuation of 175dB/km at 0.48µm is reported. Finally, in the last section, I introduce a novel approach for fast gas detection in the near infrared regime with large bandwidth HC-ARFs, which further proves the versatility of this technological approach.

## II. Design Flexibility

Figure 1 shows the fiber structures that I have used for my numerical simulations with Comsol.

All fiber structures have a core diameter $D$ of 43μm. The diameters of the cladding tubes are d=D, $d_1$=d/2, $d_2$=$d_1$/2, $d_{fbf}$=33μm, $d_{1fbf}$=$d_{fbf}$/2.

*A. Design of large bandwidth hollow core fibers*

The glass thickness $t$ of the cladding tubes is the same $t$=0.3μm for all the known structures on the left hand side and in the middle of Fig. 1: the Single Anti-Resonant (SAR) structure [as in 10], the Double Anti-resonant (DAR) structure (as in [15] and [17]), the Triple Anti-Resonant Structure (TAR, as in [15] and [17]) and the "Free core Boundary Fiber" (i.e. a fiber of the type described in [14]) with a Double Anti-Resonant Structure (DARfbf) (as in [16] and [18]). These four structures have already been taken into considerations previously but those studies concerned only the optimization of their attenuation level (confinement, bending or scattering loss). In the present study, I have adopted a very thin silica thickness of the core boundary ($t$=0.3μm) in order to maximize the transmission bandwidth of this fiber type, in the visible and near-infrared spectral window. Indeed, in HC-ARFs the wavelengths $\lambda_\kappa$ of maximum optical transmission ("antiresonance wavelengths") are given by the formula [19]:

$$\lambda_k = 4t\frac{\sqrt{n^2-1}}{(2k+1)} \qquad k=0,1,2,... \quad (1)$$

where $n$ is the glass refractive index, $t$ is the glass thickness and k is an integer which defines the order of the considered transmission window around λ (k=0 corresponds to the first antiresonant window, k=1 to the second antiresonance window and so on). The first antiresonance window is the one that allows a larger transmission bandwidth in all sort of antiresonant fibers [20]. Therefore, my choice of a thin silica core boundary $t$=0.3μm sets the first antiresonance transmission window to be located in the visible and near-infrared regime (see Fig. 2). Note that a large transmission bandwidth is very useful for applications in femtosecond pulse delivery, Raman spectroscopy or gas-based nonlinear optics.

Fig. 2 shows the leakage losses of the simulated fibers (the SAR in black, the DAR in green, the TAR in red and the DARfbf in blue).

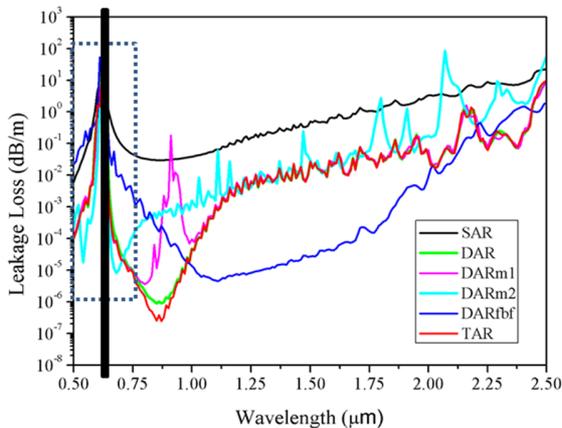
Fig. 2. Transmission bandwidth of the HC-ARFs described in Fig. 1.

The glass thickness $t$ (common to all the considered structures) set the lower limit of the bandwidth to the resonant wavelength λ=0.63μm (vertical black line). From previous works [21]-[23], a useful level to define the transmission bandwidth is set to the attenuation of 1dB/m, which can be still acceptable for some of the above cited applications. When comparing the SAR, DAR, TAR and DARfbf, we notice that their bandwidth extends between 0.65μm and 2.5μm. The maximum theoretical bandwidth of other designs of hollow core fibers in the visible or near infrared was only extending in the range of wavelengths between 1 and 2μm [22].

Although their overall transmitting bandwidth is the same, the considered structures show different transmitting spectra. In particular, as expected, the DAR and TAR structure show a reduced attenuation [17]. However the attenuation is below 1dB/km only between 0.65 μm and 1.12μm. At longer wavelength the attenuation increases (showing fast oscillations) due to the coupling between the silica cladding modes and the fundamental-like core mode. This aspect has already been treated in [11], [14] and [17]. The adoption of a "free" core boundary (in which the cladding tubes are separated by each other) presents the advantage of reducing the coupling between cladding and fundamental modes, as explained in [14]. That's why the use of the DARfbf design (blue line) shows the advantage of having a leakage loss of less than 1dB/km over a larger bandwidth (comprised between about 0.8 and 1.85μm). As explained in [14] this comes at the expense of an increased minimum leakage loss as compared to the DAR fiber.

The Group Velocity Dispersion (GVD) of the DARfbf structure is shown in Fig. 3(B). In previous designs of wide bandwidth hollow core fibers the GVD value was varying between 1ps/nm-km at about 1.15μm and 100ps/nm-km at about 1.85μm. In Fig. 3(B) we can observe a much lower variation of the GVD value between -1 ps/nm-km at 0.65μm and about 4ps/nm-km at 2.5μm. The fiber dispersion is anomalous for wavelength longer than 0.74μm.

*B. Novel designs for hollow antiresonant fibers*

In order to show the design flexibility of hollow antiresonant fibers, I have considered in Fig. 1 novel fiber structures (those on the right hand side). They are a variation of the DAR structure and I have explored them in order to increase the overall fiber transmission bandwidth, by trying to decrease the leakage loss around the resonant zone (i.e. around λ=0.63μm). The novel idea is using 2 different thicknesses for the antiresonant elements present in the fiber cladding area: the first thickness is still set to $t$=0.3μm, as in the previous designs, while the other thickness is different. Indeed one may think that matching the resonant wavelength (λ=0.63μm) of some elements with the antiresonant wavelengths of some other elements in the cladding may have beneficial effects on the leakage loss around the resonant wavelength area. More specifically, in the first Modification ("DARm1" in Fig. 1) the glass thickness of the larger cladding tubes ($t_{1A}$=$t_{1B}$) is different from the one of the inner smaller tubes ($t_{2A}$=$t_{2B}$).

Indeed $t_{1A}=t_{1B}=t=0.3\mu m$ while $t_{2A}=t_{2B}=0.45\mu m$. This sets the thickness of the inner tubes $t_2$ such that the second of its antiresonance wavelengths $\lambda_{2AR}$ matches exactly the first resonance wavelength $\lambda_{1R}$ of the outer tubes with glass thickness $t$ ($\lambda_{1R}=\lambda_{2AR}=0.63\mu m$, for a calculation of these wavelength see [19] and Eq. 1 above). In the second Modification of the original DAR structure ("DARm2" in Fig. 1), the HC-ARF is built by alternating, around the core, silica tubes with different glass thickness ($t_{1A}=t_{2A}=t=0.3\mu m$ and $t_{1B}=t_{2B}=0.196\mu m$), such that the resonant wavelength $\lambda_{1R}=0.63\mu m$ falls within the first antiresonance transmission window originated by the glass tubes with thickness $t_{1B}=t_{2B}$.

The effect of these modifications is visible in Fig. 2 and in particular in Fig. 3(A), by comparing the curves in green (DAR), purple (DARm1) and sky blue (DARm2).

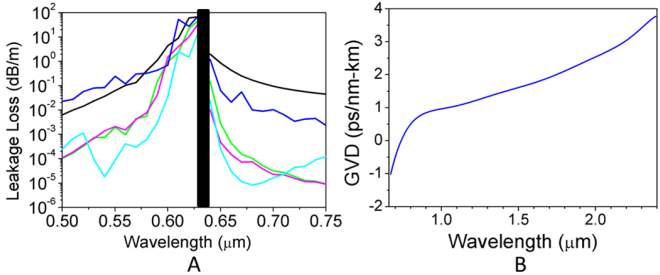

Fig. 3. On the left hand side (A) is a magnification of Fig. 2 (the area limited by the dashed points). The legend for the different colors is the same than for Fig. 2. On the right hand side (B) is the calculation of the Group Velocity Dispersion (GVD) for the structure DARfbf (represented by a blue line in Fig. 2 and 3A).

The novel introduced concept seems valuable especially when looking at the DARm2 structure (sky blue line), which allows the best reduction of leakage loss around the resonant zone (black vertical line). However this comes at the expenses of intense oscillations in the leakage loss at longer wavelengths (see Fig. 2), probably related to a stronger coupling between the fundamental mode and an increasing number of cladding modes (associated with the several silica elements with different thickness present in the fiber).

In contrast the DARm1 structure (purple line) improves only slightly the leakage loss around the resonant wavelength (black vertical line) and presents a leakage loss peak around $0.9\mu m$ (Fig. 2), which is linked to the first resonant wavelength of the glass thickness $t_{2A}=t_{2B}=0.45\mu m$.

The study of these modified structures is interesting because their development may find some use, for example in the field of spectral filtering, when one needs a high transmission extinction ratio close to the wavelength to filter out.

Although the actual improvements in the transmission bandwidth proved by these fiber modifications appear to be quite limited in the specific case considered here, these novel structures validate the idea that the use of different types of antiresonant elements within the same fiber offers an extra degree of flexibility, which can be adopted for getting an improved control of the fiber properties. I plan to explore these aspects in more detail and exploit this technique to design HC-ARFs with high birefringence, controlled modal shape, controlled dispersion, enhanced gas sensing or ultra-low attenuation.

### C. Extremely large bandwidth in the mid-infrared

Figure 4 shows that, by doubling the size of the DARfbf structure, so that the core size is now $D'=86\mu m$ and the glass thickness is $t'=0.6\mu m$, we obtain optical transmission (<1dB/m) between $1.3\mu m$ and $4.5\mu m$ (black dotted line). The contribution of the leakage loss to the total fiber attenuation is shown with a blue line, while the contribution of F300 silica attenuation [24, 25] is shown with a red line. For wavelengths lower than about $2.4\mu m$, silica F300 attenuation rapidly decreases below 0.1dB/m [24] and its contribution to the total loss of the fiber becomes negligible. The adoption of large bandwidth hollow core fibers of this type in the mid-infrared may allow the detection of a full range of gas species within the same hollow antiresonant fiber.

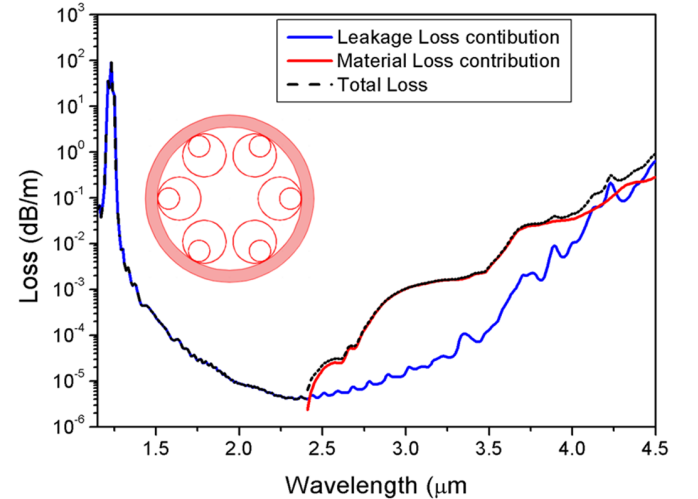

Fig. 4. Spectral attenuation of a DARfbf structure with a core size of $86\mu m$ and a glass thickness of the cladding tubes of $0.6\mu m$. All geometrical parameters of the fiber considered here are double of those relatives to the same structure considered in Fig. 2.

## III. FABRICATION AND PROPERTIES

Some of the proposed designs for HC-ARFs have been investigated experimentally.

### A. Large Bandwidth in the near infrared wavelength regime

Up to date all fabricated HC-ARFs have been reported to work in the second or in higher order antiresonance windows [10], [12]-[14]. This has been partially related to the difficulty of realizing the correct fiber structure by adopting the small glass thickness required for operation in the first antiresonance spectral window. The fiber shown in Fig. 5 (A) is the first reported case of HC-ARF that transmits light in the first window of antiresonance.

The core size of the fabricated fiber (Fig.5, A) is $23\mu m$ and the silica tickness of the core boundary is $0.34\pm0.02\mu m$. The transmission spectrum of the fibre (shown with a blue line in Fig 5) is taken by using 2m of fiber, a tungesten halogen bulb as broadband spectral source and an Optical Spectrum Analyser.

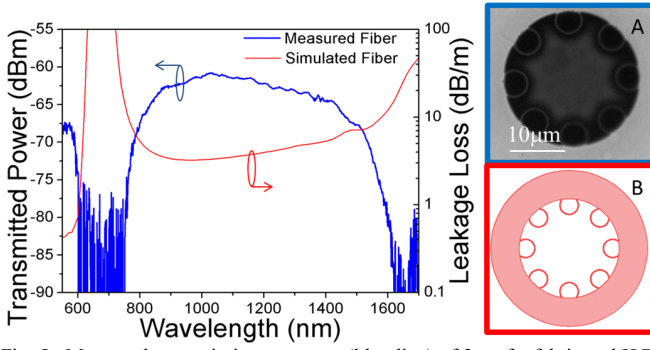

Fig. 5. Measured transmission spectrum (blue line) of 2m of a fabricated HC-ARF, working in the first antiresonance window. An optical image of the fabricated fiber, illuminated from the back, is shown on the right hand side (A). The red line shows the numerical calculation of the leakage loss of an equivalent fiber design shown on the right hand side (B).

The transmission spectrum is spanning from 750nm to 1600nm, and beats the performances of all reported HC-PBGFs [23]. A comaparable extension of the bandwidth has been obtained only in Kagome Fibers [6], but in that case the spectrum quality was affected by several optical transmission peaks along the bandwidth. On the contrary, the smoothness of the large optical spectrum reported here is related to the use of a "Free Boundary Fibre", which allows a strong reduction of the optical coupling between the fundamental and cladding modes. Fig 5 shows also the numerical calculation of the leakage loss (red line) for an equivalent fibre design (Fig. 5, B), in which the geometrical parameters of the fabricated fiber (A) have been adopted. As we can see there is a very good agreement between the experimental and numerical data in terms of transmission bandwidth.

The fabrication of a similar HC-ARF with nested cladding tubes (DARfbf in Fig. 1), working in the first antiresonant window, should allow for further increase of the transmission bandwidth.

### B. Low attenuation in the visible spectral range

The novel design for HC-ARF proposed in [15]-[17] can be interesting also for operation in higher order transmission windows, which impose less constraints on the fabrication. Some prototypes of DAR structures with a free core boundary have already been reported in [16]. Here I report about the first optical characterization of one of these fibers in the visible spectral range. All previous experimental demonstrations of all types of HC-ARFs have always concerned the near and mid-infrared wavelength regime [10], [12]-[14].

The fabricated antiresonant fiber shown on the top of Fig. 6 (A) [16] has a core diameter of 51μm and an average core boundary thickness $t_1$ of about 1.27±0.06μm. The attenuation measurement has been made by cutting a length of 14.9m of the fiber from a starting length of 20m to a final length of 5.1m. The attenuation spectrum of the fiber is on the left hand side of Fig. 6 (blue line on the top), showing a minimum attenuation of 175dB/km at λ=480nm (with an estimated maximum error of the measurement that was less than 5%). Comparable attenuations of 150dB/km at a slightly longer wavelength of 515nm (and 70dB/km at 600nm) have been reported only very recently in KFs [7]. This fiber further shows that HC-ARFs can be a valid alternative for applications in the visible spectral range.

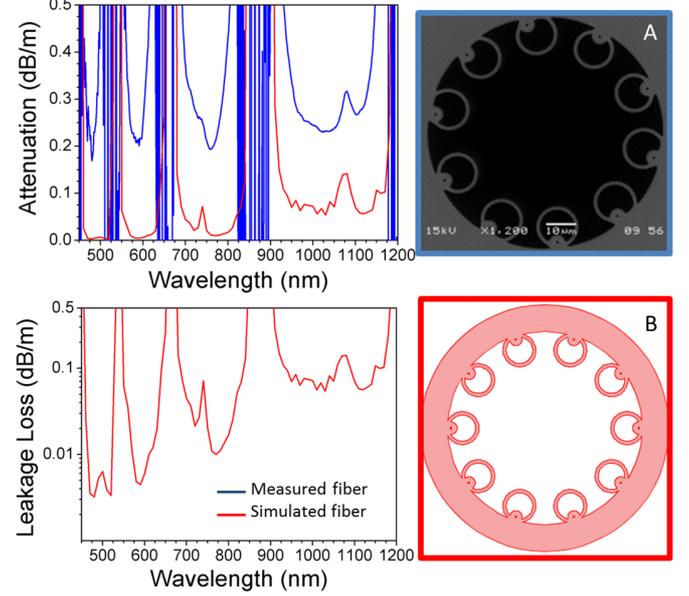

Fig. 6. Top: attenuation spectrum (blue line) of the HC-ARF shown on the right hand side. The minimum loss is 175dB/km at 480nm. Bottom: numerical calculation of the leakage loss (red line) of an equivalent fiber design (B).

The red line in Fig. 6 (in a linear scale at the top and a logarithmic scale at the bottom) shows the numerical calculation of the leakage loss of the equivalent fiber design shown on the bottom right hand side (B), in which the geometrical parameters extracted by the SEM (Fig. 6, A) have been adopted. There is a good agreement between the numerical and experimental data. Overall the large difference (particularly at shorter wavelengths) of the total attenuation of the fabricated fiber (blue line) and the leakage loss of the simulated fiber (red line) is related not only to the scattering loss (not taken into account in the simulations) but also to the several imperfections in the fabricated fiber transversal structure and to the fiber uniformity along its measured length. Further investigations will better address this loss mismatch.

The SEM of the fabricated fiber (Fig. 6, A) shows some limits that currently exist in the fabrication of this fiber. The small holes within the larger cladding tubes are not completely open and the thickness $t_2$ of the glass surrounding them is of 2.16±0.06μm, much larger than $t_1$. Due to their small size the effect of the nested ring on the fiber performances is limited. Their effect is visible from the simulation on the bottom of Fig. 6 where the loss peaks at about 500nm, 750nm and 1100nm corresponds to resonances induced by the thickness $t_2$ of the most inner holes. In the actual measurement of the fiber loss (top of Fig. 6), this effect is clearly visible only at about 1100nm where the contribution of the leakage loss to the total attenuation of the fiber is more relevant. I should also note that the minimum attenuation wavelength λ (480nm) corresponds exactly to the 10[th] order antiresonance wavelength of the glass thickness $t_2$ (from Eq. 1). Future improvements in the fabrication of this type of fiber is likely to show further loss reduction in the visible as well as in the ultra-violet spectral range.

*C. Fabrication tolerances*

Another problem which is visible from the SEM in Fig. 6 and Fig. 7 is the angle between the fiber transversal axis and the touching point between the smaller and larger cladding tubes (see Fig. 8, angle "α"). This is related to imperfections during the fiber preform preparation. The technique that I have used to fabricate the first prototypes of the DAR structure consists in a modification of the original "stack and draw" technique used for the fabrication of HC-PBGFs [2]. In a first fabrication stage one silica tube is fixed inside a second silica tube with the same wall thickness and double external diameter, and then drawn to a preliminary silica cane with an external diameter of about 3 mm. In the second fabrication stage, an assembly ("stack") of the obtained canes is made, within another jacket silica tube, to form the desired structure, and then drawn to a final silica cane. In the third and final fabrication stage, the final cane is drawn for reducing its dimensions to those of an optical fiber with a diameter of 100 to few hundreds μm. During the last stage a differential pressure is applied both to the core and to the cladding tubes for preserving the cladding structure.

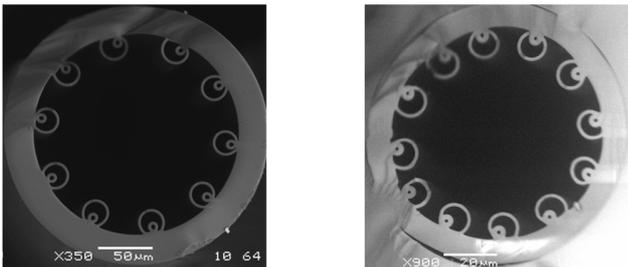

Fig. 7. SEM of two prototypes of HC-ARF with a DARfbf structure. The fiber on the left hand side was optimized for guidance in the mid-infrared [16] while the fiber on the right hand side is optimized for the near-infrared spectral range. Both fibers are affected by imperfections in the stacking of the inner tubes.

While the collapse of the smaller holes within the larger cladding tubes depends on the pressures adopted during the final fiber fabrication stage, the presence of an angle α≠0 is originated by the first and second fabrication stage. In order to show the impact that these imperfections have on the fiber performances, I have used the same ideal fiber structure already thoroughly studied in [17]. Several fiber structures are shown in Fig. 8 where the angle α varies for all tubes between 0° and 102°. The calculation of the leakage loss is made, as in [17], for the best geometrical parameters and at a wavelength of 3.05μm. Fig. 8 shows that when the angle α varies from 0° to about 90° the level of leakage loss stays below $10^{-4}$ dB/m. For angles higher than about 90° the leakage loss starts to increase. This may be related to the fact that the contact point between the internal and external tubes moves closer to the fiber core, increasing the coupling between cladding modes and fundamental-like mode [11], [14]. These numerical results show us that very large fabrication tolerances can be allowed during the preform stacking preparation of this fiber type.

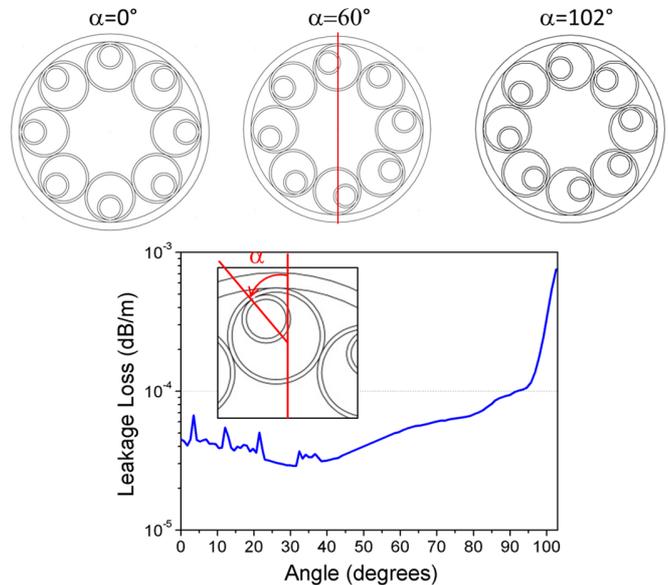

Fig. 8. Fabrication tolerances for HC-ARFs with nested cladding tubes. For the specific design considered here [17] the angle α can be varied up to about 90° with a very limited incidence on the fiber optical attenuation.

## IV. NOVEL DESIGN FOR GAS SENSING

The versatility of HC-ARFs can be further exploited for adapting them to a set of different applications. In particular HC-ARFs with a large transmission bandwidth in the near infrared spectral range can also be useful for optical detection of different gas species. A modification of the design of a HC-ARF with a free core boundary [14] can be of high interest for its applications in optical sensing or quantum information.

Fig. 9 shows the design of an HC-ARF with a free core boundary in which a lateral cut has been applied on the external fiber jacket (on the top). Lateral drilling by fs laser sources has already been investigated in HC-PBGFs for fast-responding gas detection (methane, $CH_4$), in the region between 1.5 and 1.7μm [26]. However lateral drilling involves breaking the internal structures of HC-PBGFs, and therefore originates additional optical attenuation. This limits the length of the lateral slices that can be practically processed on the fiber side, and consequently limits the possible response time of the gas sensor devices.

The design proposed in Fig. 9 does not suffer this limitation. A lateral cut of the external jacket does not involve any modification of the internal optical core boundary of the fiber. Therefore no extra guiding loss would be generated by a lateral cut, allowing the drilling of long or multiple laterals slices, and fast access to the environmental agent to be detected.

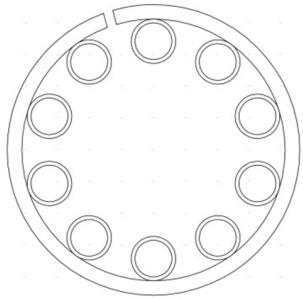

Fig. 9. Novel design for HC-ARF with a free core boundary. A lateral cut is applied on the side of the fiber (on the top) to allow the filling of an environmental agent within the fiber. No additional optical attenuation is induced by the lateral drilling of the fiber.

In order to provide a numerical demonstration of this effect, I have used the DARfbf fiber design (see Fig. 1) already simulated in Fig. 2 (blue line). I have then applied a lateral cut to this fiber design, as shown in Fig. 10 (B). I have performed a new numerical simulation on this modified design (B) by adopting exactly the same model parameters used for the original design (A). The results are shown in Fig. 10, between 0.8 and 1.85μm. As we can see, the application of a lateral cut to the original fiber design doesn't cause any relevant increase in the fiber attenuation. The additional loss is limited to an amount in the order of 0.01dB/km only, which I believe to be related to the different contribution of the Fresnel reflections at the air-glass interface of the fiber jacket.

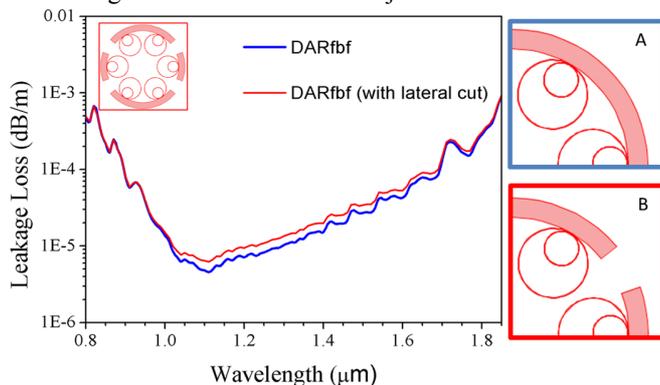

Fig. 10. Comparison between the leakage loss of an antiresonant fiber (fiber A, blue line) and the same fiber with a lateral cut on the fiber jacket (fiber B, red line). The numerical calculation is performed on one quarter of the fiber structure: the inset on the top shows the complete structure of fiber B.

Note that the numerical simulations have taken into account only a quarter of the fiber structure (Fig. 10, right hand side), in order to improve the accuracy of the numerical analysis. This means that the results of the simulations refer to the full fiber design shown in the inset of Fig. 10, which present 4 lateral cuts (due to symmetry rules [27]).

The lateral drilling of hollow core antiresonant fibers maybe employed also in the implementation of optical memories with high optical depth, which would require a fast filling of HC-ARFs with Caesium vapors [28].

## V. CONCLUSION

In this work I have discussed some of the new possibilities offered by different designs of HC-ARFs, in the near-infrared and visible spectral range.

I have shown that specific structures of HC-ARFs can have very large transmission bandwidth in the near infrared, much larger than those achievable with HC-PBGFs. A design for a HC-ARF with a bandwidth spanning from 0.65μm to 2.5μm has been reported. The fabrication of a HC-ARF with a transmission bandwidth between 0.75μm and 1.6μm has also been reported. This HC-ARF was working, for the first time, in the first antiresonance transmission window.

I have reported on the optical characterization of a HC-ARF with a very low attenuation of 175dB/km at 0.48μm. I have discussed the fabrication issues of this fiber type and some relevant fabrication tolerances.

Finally I have proposed that some of the fiber structures of HC-ARFs, with large transmission bandwidth in the near infrared spectral range, could be also used for allowing fast gas detection. The lateral cutting of HC-ARFs with a free core boundary may be employed for enhancing the performances of distributed optical sensors or optical memories.

ACKNOWLEDGMENT

I have fabricated and characterized the reported fibers while working at the University of Bath. I would like to thank Jonathan Knight for the great support for this research activity, for motivation and several scientific discussions. I would like to thank Peter Mosley for suggesting me that the lateral cutting of HC-ARFs may have been also employed in the field of quantum information. I thank Francesco Poletti and David Richardson for useful discussions during the preparation of this manuscript.

REFERENCES

[1]  E. Marcatili and R. Schmeltzer, "Hollow metallic and dielectric waveguides for long distance optical transmission and lasers", Bell Syst. Tech. J. 43, 1783-1809 (1964).
[2]  R. F. Cregan, B. J. Mangan, J. C. Knight, T. A. Birks, and P. St. J. Russell," Single-Mode Photonic Band Gap Guidance of Light in Air" Science 285, 1537 (1999).
[3]  P. J. Roberts, F. Couny, H. Sabert, B. J. Mangan, D. P. Williams, L. Farr, M. W. Mason, A. Tomlinson, T. A. Birks, J. C. Knight and P. St. J. Russell, "Ultimate low loss of hollow-core photonic crystal fibres," Opt. Express 13, 236-244 (2005).
[4]  NKTphotonics, *http://www.nktphotonics.com/*.
[5]  M. A. Duguay, Y. Kokubun, T. L. Koch, and L. Pfeiffer, "Antiresonant reflecting optical waveguides in $SiO_2$Si multilayer structures," Appl. Phys. Lett. 49(1), 13-15 (1986).
[6]  Y. Y. Wang, N. V. Wheeler, F. Couny, P. J. Roberts, and F. Benabid, "Low loss broadband transmission in hypocycloid-core Kagome hollow-core photonic crystal fiber," Opt. Lett. 36(5), 669–671 (2011).
[7]  B. Debord, M. Alharbi, A. Benoît, D. Ghosh, M. Dontabactouny, L. Vincetti, J.-M. Blondy, F. Gérôme, and F. Benabid, "Ultra low-loss hypocycloid-core Kagome hollow-core photonic crystal fiber for green spectral-range applications," Opt. Lett. 39, 6245-6248 (2014).
[8]  G. J. Pearce, G. S. Wiederhecker, C. G. Poulton, S. Burger, and P. St. J. Russell, "Models for guidance in kagome-structured hollow-core photonic crystal fibres," Opt. Express 15, 12680-12685 (2007).
[9]  S. Février, B. Beaudou, and P. Viale, "Understanding origin of loss in large pitch hollow-core photonic crystal fibers and their design simplification," Opt. Express 18(5), 5142–5150 (2010).
[10] A. D. Pryamikov, A. S. Biriukov, A. F. Kosolapov, V. G. Plotnichenko, S. L. Semjonov, and E. M. Dianov, "Demonstration of a waveguide regime for a silica hollow-core microstructured optical fiber with a

**Walter Belardi** obtained his Laurea degree in Electronics Engineering from the University of Pavia, Italy, and his Ph.D degree in Optoelectronics from the Optoelectronics Research Centre, University of Southampton.
He was a postdoctoral researcher with the Université des Sciences et Technologies de Lille, France. Before coming back to Academia, he worked in the private sector, as research engineer in STMicroelectronics and as scientific consultant in a spin-off company of the University of Lille. He re-joined the Optoelectronics Research Centre in Southampton (UK) in September 2014, after having worked at the University of Bath (UK) within the EPSRC funded project: "New Fibers for new lasers: photonic crystal fibre optics for the delivery of high-power light".

His principal research contributions are in the design, fabrication and use of novel optical fibre technologies. His most important personal achievements range from the first application of the spinning technique to Microstructured Optical Fibers (MOFs), the first inclusion of an elliptical hole within a MOF, the invention of a double air clad fabrication approach for MOFs, to the most recent results on hollow core optical fibres. These include theoretical studies on their geometrical structure, the introduction of a radically novel design for low loss hollow core optical fibers and the first theoretical and experimental demonstration of low bending loss in hollow antiresonant fibers.
More details about his research are on his blog: www.walterbelardi.com